\documentclass[conference,10pt]{IEEEtran}
\usepackage{epsfig,rotating,setspace,latexsym,amsmath,epsf,amssymb,amsfonts,bm,theorem,subfigure,epstopdf}
\usepackage{cite,authblk}
\usepackage{bbm}
\usepackage{algorithm}
\usepackage[noend]{algpseudocode}

\algrenewcommand\algorithmicforall{\textbf{foreach}}
\algrenewcommand\algorithmicindent{.8em}
\usepackage{color}
\usepackage{mathtools}

\newtheorem{theorem}{Theorem}
\newtheorem{lemma}{Lemma}

\newenvironment{Proof}[1]{\medskip\par\noindent{\bf Proof:\,}\,#1}{{\mbox{\,$\blacksquare$}\par}}
\IEEEoverridecommandlockouts
\allowdisplaybreaks

\begin{document}

\title{Optimal Selective Encoding for Timely Updates  \thanks{This work was supported by NSF Grants CCF 17-13977 and ECCS 18-07348. }}
\author{Melih Bastopcu \qquad Baturalp Buyukates  \qquad Sennur Ulukus\\
	\normalsize Department of Electrical and Computer Engineering\\
	\normalsize University of Maryland, College Park, MD 20742\\
	\normalsize  \emph{bastopcu@umd.edu}\hspace{0.5em} \qquad \emph{baturalp@umd.edu}  \qquad \hspace{0.5em} \emph{ulukus@umd.edu}}
\maketitle

\begin{abstract}	
 We consider a system in which an information source generates independent and identically distributed status update packets from an observed phenomenon that takes $n$ possible values based on a given pmf. These update packets are encoded at the transmitter node to be sent to the receiver node. Instead of encoding all $n$ possible realizations, the transmitter node only encodes the most probable $k$ realizations and disregards whenever a realization from the remaining $n-k$ values occurs. We find the average age and determine the age-optimal real codeword lengths such that the average age at the receiver node is minimized. Through numerical evaluations for arbitrary pmfs, we show that this selective encoding policy results in a lower average age than encoding every realization and find the age-optimal $k$. We also analyze a randomized selective encoding policy in which the remaining $n-k$ realizations are encoded and sent with a certain probability to further inform the receiver at the expense of longer codewords for the selected $k$ realizations. 
\end{abstract}
 
\section{Introduction}
Age of information is a performance metric which quantifies the timeliness of information in networks. It keeps track of the time since the most recent update at the receiver has been generated at the transmitter. Age increases linearly in time such that at time $t$ age $\Delta(t)$ of an update packet which was generated at time $u(t)$ is $\Delta(t) = t-u(t)$. When a new update packet is received, the age drops to a smaller value. Although initial applications of age of information considered  queueing networks, scheduling and optimization problems \cite{Kaul12a, Costa14, Bedewy16, He16a, Sun17a, Najm18b, Najm17, Soysal18, Soysal19, Yates17b, Hsu18b, Kadota18a, Gong19, Buyukates19c, Arafa19b, Sun17b, Sun18b, Bastopcu19, partial_updates, Zou19b, Non_linear, bastopcu_soft_updates_journal, Arafa17b, Arafa17a, Wu18,Arafa_Age_Online, Arafa18f, Arafa19e, Farazi18, Yener_energy_19}, the concept of age is applicable to a wider range of problems, particularly in autonomous driving, augmented reality, social networks, and online gaming, as information freshness is crucial in all these emerging applications.

\begin{figure}[t]
	\centering  \includegraphics[width=1\columnwidth]{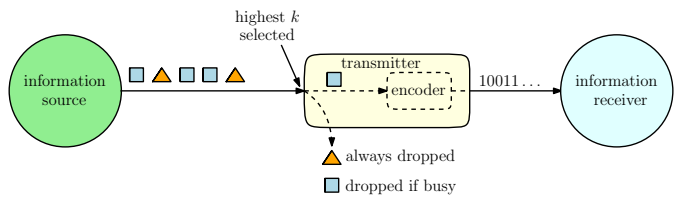}
	\caption{An information source generates i.i.d. status updates from a random variable $X$. Only a portion of the realizations (shown with a square) is encoded into codewords. Update packets that come from the selected portion of the realizations that find the transmitter node idle are sent to the receiver node. Non-selected realizations (shown with a triangle) are always discarded at the transmitter node even if the transmitter node is idle. }
	\label{fig:model}
	\vspace{-0.5cm}
\end{figure}

In this work, we consider a status updating system that consists of a single transmitter node and a single receiver node (see Fig.~\ref{fig:model}). The transmitter node sends independent and identically distributed status update packets generated by an information source regarding an observed random phenomenon which takes finitely many values based on a known pmf. Unlike most of the existing literature, we focus on the \emph{source coding} aspect of the problem and design codewords to minimize the average age experienced by the receiver node. 

References that are most closely related to our work are \cite{Mayekar18, Zhong16, Yates_Soljanin_source_coding} which study the timely source coding problem. Reference \cite{Mayekar18} considers a zero-wait policy and finds codeword lengths which achieve the minimum average age up to a constant gap at the receiver node by using Shannon codes based on a modified version of the given pmf. References \cite{Zhong16} and \cite{Yates_Soljanin_source_coding} study the block coding and source coding problems, respectively, for a queueing theoretic setting and consider the age-optimal codes for a FIFO queue.

Inspired by these works, we consider a \emph{selective encoding} mechanism at the transmitter node. In other words, instead of encoding all possible realizations, in our model, we encode only a portion of the realizations and send to the receiver node. Specifically, we consider what we call the \emph{highest $k$ selective encoding} scheme in which we only encode the most probable $k$ realizations and disregard any update packets from the remaining $n-k$ realizations. Similar $k$ out of $n$ type of schemes are shown to achieve better age performance especially in the context of multicast networks in which each update packet is transmitted until the earliest $k$ of the $n$ receiver nodes receive that packet \cite{Zhong17a, Zhong18b, Buyukates18, Buyukates18b, Buyukates19}. We note that a smaller $k$ in our setting yields shorter codeword lengths but larger interarrival times, as in this case most of the updates are not encoded. However, when $k$ is large, codeword lengths and correspondingly the transmission times get larger even though the interarrival times get smaller. Thus, in this paper, based on the given pmf, we aim to find the optimal $k$ which strikes a balance between these two opposing trends such that the average age at the receiver node is minimized. Due to this selective encoding scheme not every realization is sent to the receiver even if the channel is free.

In this work, we find the average age expression under the highest $k$ selective encoding scheme and determine the age-optimal real codeword lengths such that the average age at the receiver is minimized. For arbitrary pmfs, we numerically show that this selective encoding policy achieves a lower age than encoding all the realizations and find the age-optimal $k$. Lastly, we consider a scenario in which the remaining $n-k$ realizations are not completely disregarded but encoded with a certain probability which we call the \emph{randomized selective encoding} scheme. We derive the average age for this case and determine the age-optimal real codeword lengths.
                    
\section{System Model and Problem Formulation}\label{sect:model}
We consider a communication system in which an information source generates independent and identically distributed status update packets from an observed phenomenon that takes realizations from the set $\mathcal{X}= \{x_1, x_2, \ldots, x_n\}$ based on a known pmf $P_X(x_i)$ for $i \in \{1, \ldots, n\}$. Without loss of generality, we assume that $P_X(x_m) \geq P_X(x_j)$ for all $m\leq j$, i.e., the probabilities of the realizations are in a non-increasing order. Update packets arrive at the transmitter node following a Poisson process with parameter $\lambda$. The transmitter node implements a blocking policy in which the update packets that arrive when the transmitter node is busy are blocked and lost. Thus, the transmitter node receives only the updates which arrive when it is idle. 

In this paper, unlike \cite{Mayekar18} which considers deterministic or randomized encoding, we consider a selective encoding mechanism that we call \emph{highest k selective encoding} where the transmitter node only sends the most probable $k$ realizations, i.e., only the realizations from set $\mathcal{X}_k =\{x_1, \ldots, x_k\}$, which have the highest probabilities among possible $n$ updates generated by the source, are transmitted for $k \in \{1,\ldots, n\}$.\footnote{In fact, instead of the most probable $k$ realizations, the transmitter can select any $k$ of $n$ possible realizations for encoding to assure that a pre-selected group of realizations is always encoded.} If an update packet from the remaining non-selected portion of the realizations arrives, the transmitter disregards that update packet and waits for the next arrival. If an update packet arrives from the selected portion of the realizations, then the transmitter encodes that update packet by using a binary alphabet with the conditional probabilities given by, 
\begin{align}\label{cond_prob}
   P_{X_k}(x_i)= \begin{cases} 
      \frac{P_X(x_i)}{q_k}, & i=1,2,\ldots, k \\
      0, & i=k+1,k+2,\ldots, n,   
   \end{cases}
\end{align}
where $q_k \triangleq \sum_{\ell=1}^{k}P_X(x_\ell)$. The transmitter assigns codeword $c(x_i)$ with length $\ell(x_i)$ to realization $x_i$ for $i\in\{1,2,\ldots,k\}$. The first and the second moments of the codeword lengths are given by $\mathbb{E}[L] = \sum_{i=1}^{k}P_{X_k}(x_i)\ell(x_i)$ and $\mathbb{E}[L^2] = \sum_{i=1}^{k}P_{X_k}(x_i)\ell(x_i)^2$. The channel between the transmitter node and the receiver node is assumed to be error-free. The transmitter node sends one bit at a unit time. Thus, if the transmitter node sends update $x_i$ to the receiver node, this transmission takes $\ell(x_i)$ units of time. That is, for realization $x_i$, the service time of the system is $\ell(x_i)$.

In this paper, we consider the age metric to measure the freshness of the information at the receiver node. Let $\Delta(t)$ be the instantaneous age at the receiver at time $t$ with $\Delta(0)= \Delta_0$. Age at the receiver node increases linearly in time and drops to the age of the most recently received update when a delivery occurs. We define the long term average age as,
\begin{align}
    \Delta = \lim_{T\to\infty} \frac{1}{T}\int_0^T\Delta(t)dt.
\end{align}

Our aim is to find the codeword lengths that minimize the average age for a given $k$ such that a uniquely decodable code can be designed, i.e., the Kraft inequality is satisfied \cite{Cover}. Thus, we formulate the problem as,
\begin{align}
\label{problem1}
\min_{\{ \ell(x_i) \}}  \quad & \Delta \nonumber \\
\mbox{s.t.} \quad & \sum_{i=1}^{k} 2^{-\ell(x_i)}\leq 1 \nonumber \\
\quad & \ell(x_i) \in \mathbb{Z}^+, \quad i\in\{1,\ldots,k\}.
\end{align}
In the following section, we find an analytical expression for the long term average age, $\Delta$. 

\section{Average Age Analysis}\label{sect:age_analysis}
As described in Section~\ref{sect:model}, status update packets arrive at the transmitter node as a Poisson process with rate $\lambda$. The transmitter node implements a blocking policy in which update packets that arrive when the transmitter node is busy are blocked from entry and dropped. Thus, upon successful delivery of a packet to the receiver node, the transmitter idles until the next update packet arrives. This idle waiting period in between two arrivals is denoted by $Z$ which is an exponential random variable with rate $\lambda$ as update interarrivals at the transmitter node are exponential with $\lambda$. Not every packet which enters the transmitter node is sent to the receiver node. The transmitter node implements the highest $k$ selective encoding policy in which only the most probable $k$ realizations are encoded. When one of the remaining $n-k$ packets enters the transmitter node, however, transmitter node continues to wait for the next update arrival. 

We denote the update packets which belong to the set $\mathcal{X}_k$ and arrive when the transmitter is idle as the successful update packets. Since the channel is error-free and there is no preemption, these successful packets are received by the receiver node. Let $T_{j-1}$ denote the time instant at which the $j$th successful update packet is received. We define the time in between two successive successful update arrivals at the transmitter node as the update cycle and denote it with $Y_j = T_j - T_{j-1}$. Update cycle $Y_j$ consists of a busy cycle and an idle cycle such that,
\begin{align}
	Y_j = S_j + W_j, \label{update_cycle}
\end{align}
where $S_j$ denotes the service time of update $j$ and $W_j$ denotes the overall waiting time in the $j$th update cycle such that $W = \sum_{\ell=1}^{M}Z_{\ell}$. Here, $M$ is a geometric random variable with parameter $q_k \triangleq \sum_{\ell=1}^{k}P_X(x_{\ell})$ which denotes the total number of 
update arrivals until the first update from the set $\mathcal{X}_k$ is observed at the transmitter node. $W$ is also an exponential random variable with rate $\lambda q_k$ \cite[Prob. 9.4.1]{Yates14}. The moments of $M$ are
\begin{align}
\mathbb{E}[M] &= \frac{1}{{q_k}}, \label{moment1} \\  
\mathbb{E}[M^2] &= \frac{2-{q_k}}{q_k^2}. \label{moment2}
\end{align}

\begin{figure}[t]
	\centering  \includegraphics[width=0.85\columnwidth]{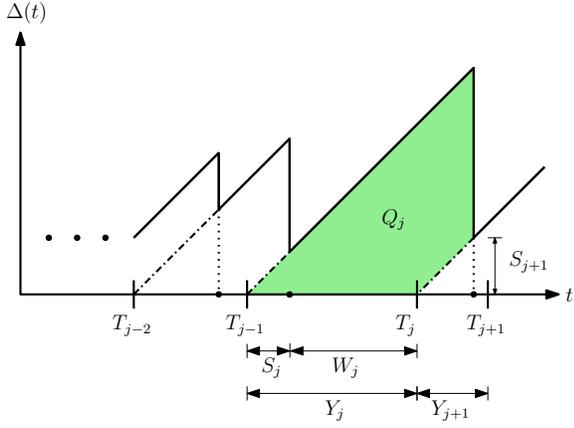}
	\caption{Sample age evolution $\Delta{(t)}$ at the receiver node. Successful updates are indexed by $j$. The $j$th successful update arrives at the server node at $T_{j-1}$. Update cycle at the server node is the time in between two successive arrivals and is equal to $Y_j = S_j + W_j = T_{j} - T_{j-1}$.}
	\label{fig:ageEvol}
	\vspace{-0.5cm}
\end{figure}

We note that arrival and service processes are independent, i.e., $S_j$ and $Z_j$ are independent, and $M$ is independent of $S_j$ and $Z_j$. Sample age evolution at the receiver node is given in Fig.~\ref{fig:ageEvol}. Here, $Q_j$ denotes the area under the instantaneous age curve in update cycle $j$ and $Y_j$ denotes the length of the $j$th update cycle as defined earlier. The metric we use, long term average age, is the average area under the age curve which is given by \cite{Najm17}
\begin{align}
	\Delta = \limsup_{n\to\infty} \frac{\frac{1}{n}\sum_{j=1}^{n}Q_j}{\frac{1}{n}\sum_{j=1}^{n}Y_j} = \frac{\mathbb{E}[Q]}{\mathbb{E}[Y]}. \label{avg_age1}
\end{align}
By using Fig.~\ref{fig:ageEvol}, we find 
\begin{align}
    Q_j = \frac{1}{2}Y^2_j + Y_j S_{j+1}, \label{shaded_area}
\end{align}
where $Y_j$ is given in (\ref{update_cycle}). Noting that $\mathbb{E}[S] = \mathbb{E}[L]$, where $\mathbb{E}[L]$ is the expected length of the codewords, (\ref{avg_age1}) is equivalent to
\begin{align}
    \Delta =  \frac{\mathbb{E}[Y^2]}{2\mathbb{E}[Y]}+\mathbb{E}[L]. \label{avg_age2}
\end{align}

Next theorem characterizes the average age, $\Delta$, given in (\ref{avg_age2}) for our model described in Section~\ref{sect:model}.
\begin{theorem}\label{thm1}
    Under the highest $k$ selective encoding scheme, the average age at the receiver node is given by
    \begin{align}
        \Delta =  \frac{\mathbb{E}[L^2] + \frac{2}{q_k\lambda} \mathbb{E}[L] + \frac{2}{(q_k\lambda)^2}}{2\left(\mathbb{E}[L] +\frac{1}{q_k\lambda}\right)}+\mathbb{E}[L], \label{avg_age3} 
    \end{align}
with $\mathbb{E}[L]$, $\mathbb{E}[L^2]$ and $q_k$ are as defined earlier.
\end{theorem}{}
\begin{Proof}
By inspecting the figure and noting that $M$, $L$ and $Z$ are mutually independent, we find \cite[Thm. 9.11]{Yates14},
\begin{align}
      \mathbb{E}[Y] =& \mathbb{E}[L] + \mathbb{E}[M]\mathbb{E}[Z], \label{E[Y]}\\
      \mathbb{E}[Y^2] =& \mathbb{E}[L^2] + 2\mathbb{E}[M]\mathbb{E}[Z]\mathbb{E}[L]+\mathbb{E}[M]\mathbb{E}[Z^2] \nonumber \\ &+(\mathbb{E}[M^2]-\mathbb{E}[M])(\mathbb{E}[Z])^2\label{E[Y^2]},
\end{align}
where moments of $M$ follow from (\ref{moment1}) and (\ref{moment2}), and $Z$ has exponential distribution with rate $\lambda$ as discussed earlier. Substituting (\ref{E[Y]}) and (\ref{E[Y^2]}) in (\ref{avg_age2}) yields the result in (\ref{avg_age3}).
\end{Proof}{}

Thus, (\ref{avg_age3}) characterizes the average age, $\Delta$, achieved at the receiver node in terms of the first and second moments of the codeword lengths for a given pmf, selected $k$, and update arrival rate $\lambda$. In the next section, we focus on age-optimal codeword design, and solve the problem in (\ref{problem1}).

\section{Optimal Codeword Design} \label{sect:opt_soln}
In this section, we find the optimal code lengths that minimize $\Delta$ given in (\ref{avg_age3}). We relax the optimization problem in (\ref{problem1}) and allow real length codewords. The modified optimization problem becomes
\begin{align}
\label{problem1_mod}
\min_{\{ \ell(x_i) \}}  \quad &  \frac{\mathbb{E}[L^2]+2a\mathbb{E}[L]+2a^2}{2(\mathbb{E}[L]+a)} +\mathbb{E}[L] \nonumber \\
\mbox{s.t.} \quad & \sum_{i=1}^{k} 2^{-\ell(x_i)}\leq 1 \nonumber \\
\quad & \ell(x_i) \in \mathbb{R}^+, \quad i\in\{ 1,\ldots, k\},
\end{align}
where $a = \frac{1}{\lambda q_k}$. Similar to \cite{Sun17b} and \cite{Arafa_Age_Online}, we define $p(\theta)$ as
\begin{align}\label{p_theta}
    p(\theta):=& \min_{\{ \ell(x_i) \}}\frac{1}{2}\mathbb{E}[L^2]+\mathbb{E}[L]^2+(2a-\theta)\mathbb{E}[L]+a^2-\theta a \nonumber\\
    &\quad \mbox{s.t.} \quad \sum_{i=1}^{k} 2^{-\ell(x_i)}\leq 1 \nonumber \\
    & \qquad \quad \hspace{0.5em} \ell(x_i) \in \mathbb{R}^+, \quad i\in\{ 1,\ldots, k\}.
\end{align}
One can show that $p(\theta) $ is decreasing in $\theta$ and the optimal solution is obtained when $p(\theta) = 0$ such that the optimal age for the problem in (\ref{problem1_mod}) is equal to $\theta$, i.e., $\Delta^*= \theta $ \cite{frac_programming}. We define the Lagrangian \cite{Boyd04} function as
\begin{align}
    \mathcal{L} =&\frac{1}{2}\mathbb{E}[L^2]+\mathbb{E}[L]^2+(2a-\theta)\mathbb{E}[L]+a^2-\theta a \nonumber \\ &+\beta\left(\sum_{i=1}^{k} 2^{-\ell(x_i)}- 1  \right),
\end{align}
where $\beta \geq 0$. Next, we write the KKT conditions as
\begin{align}\label{KKT_cond}
    \frac{\partial\mathcal{L}}{\partial \ell(x_i)} =&  P_{X_k}(x_i)\ell(x_i)+2\left(\sum_{j=1}^{k}P_{X_k}(x_j)\ell(x_j)\right)P_{X_k}(x_i) \nonumber\\
    &+ (2a-\theta)P_{X_k}(x_i)-\beta (\log2)2^{-\ell(x_i)} = 0, \quad \forall i 
\end{align}
The complementary slackness condition is
\begin{align}\label{CS_cond}
   \beta\left(\sum_{i=1}^{k} 2^{-\ell(x_i)}- 1  \right) = 0. 
\end{align}
In the following lemma, we prove that the optimal codeword lengths must satisfy the Kraft inequality as an equality. 
\begin{lemma}\label{Lemma:CS}
	For the age-optimal real codeword lengths, we must have $ \sum_{i=1}^{k} 2^{-\ell(x_i)}= 1$.
\end{lemma} 
\begin{Proof}
	Assume that the optimal codeword lengths satisfy $\sum_{i=1}^{k} 2^{-\ell(x_i)}<1$, which implies that $\beta= 0$ due to (\ref{CS_cond}). From (\ref{KKT_cond}), we have
	\begin{align}
	P_{X_k}(x_i)\ell(x_i)+2\left(\sum_{j=1}^{k}P_{X_k}(x_j)\ell(x_j)\right)P_{X_k}(x_i) \nonumber\\
    + (2a-\theta)P_{X_k}(x_i) = 0, \quad \forall i.
	\end{align}
	Then, we find $\ell(x_i)= \frac{\theta-2a}{3}$ for all $i \in \{1,2,\ldots,k\}$. Thus, $\mathbb{E}[L]= \frac{\theta-2a}{3}$ and $ \mathbb{E}[L^2]=\left( \frac{\theta-2a}{3}\right)^2$ so that $p(\theta) = -\frac{\theta^2}{6}-\frac{\theta a}{3} + \frac{a^2}{3}$. By using $p(\theta) = 0$,  we find  $\theta = (-1+\sqrt{3})a$ which gives $\ell(x_i)=\frac{(-3+\sqrt{3})a}{3} < 0$ for $i\in\{1,2,\ldots, k\} $. Since the codeword lengths cannot be negative, we reach a contradiction. Thus, the optimal codeword lengths must satisfy $\sum_{i=1}^{k} 2^{-\ell(x_i)}=1$.   
\end{Proof}

Next, we find the optimal codeword lengths which satisfy $\sum_{i=1}^{k} 2^{-\ell(x_i)}= 1$. By summing (\ref{KKT_cond}) over all $i$, we obtain
\begin{align}\label{E_L_thr}
  \mathbb{E}[L] = \frac{\theta +\beta \log2-2a}{3}.  
\end{align}
From (\ref{KKT_cond}), we obtain
 \begin{align} \label{KKT_mod}
     -\ell(x_i) +\frac{\beta\log2}{P_{X_k}(x_i)} 2^{-\ell(x_i)} = 2\mathbb{E}[L]+2a-\theta,
 \end{align}
 for $i\in\{1,2,\ldots, k\}$, which yields
 \begin{align}\label{KKT_mod2}
     \frac{\beta (\log2)^2}{P_{X_k}(x_i)}2^{-\ell(x_i)}e^{\frac{\beta (\log2)^2}{P_{X_k}(x_i)}2^{-\ell(x_i)}} = \frac{\beta (\log2)^2}{P_{X_k}(x_i)}2^{\frac{-\theta+2\beta\log2+2a}{3}}. 
 \end{align}
Note that (\ref{KKT_mod2}) is in the form of $xe^x=y$ where the solution for $x$ is equal to $x = W_0(y)$ if $y\geq 0$. Here, $W_0(\cdot)$ denotes the principle branch of the Lambert $W$ function \cite{lambert}. Since the right hand side of (\ref{KKT_mod2}) is always non-negative, we are only interested in $W_0(\cdot)$ which is denoted as $W(\cdot)$ from now on. We find the unique solution for $\ell(x_i)$ as
\begin{align}\label{eqn:opt_lengths}
\ell(x_i) =-\frac{\log \left( \frac{ P_{X_k}(x_i)}{\beta (\log2)^2} W\left(\frac{\beta (\log2)^2}{P_{X_k}(x_i)} 2^{\frac{-\theta +2\beta \log2+2a}{3}} \right)\right)}{\log2},
\end{align} 
for $i\in\{1,2,\ldots, k\}$. 

In order to find the optimal codeword lengths, we solve (\ref{eqn:opt_lengths}) for a $(\theta, \beta)$ pair that satisfies $p(\theta)  = 0$ and the Kraft inequality, i.e., $\sum_{i=1}^{k} 2^{-\ell(x_i)} = 1$. Starting from an arbitrary $(\theta, \beta)$ pair, if $p(\theta)>0$ (or $p(\theta)<0$), we increase (or respectively decrease) $\theta$ in the next iteration as $p(\theta)$ is a decreasing function of $\theta$. Then, we update $\beta$ by using (\ref{E_L_thr}). We repeat this process until $p(\theta)  = 0$ and $\sum_{i=1}^{k} 2^{-\ell(x_i)} = 1$.   

We note that the age-optimal codeword lengths found in this section are for a fixed $k$. Thus, depending on the selected $k$, different age performances are achieved at the receiver node. In Section~\ref{sect:num_res}, we find the age-optimal $k$ values for some given arbitrary pmfs numerically.

Under the highest $k$ selective encoding policy, the receiver node does not receive any update when the remaining $n-k$ realizations occur. However, there may be scenarios in which these remaining realizations are also of interest to the receiver node. In the next section, we focus on this scenario and consider a randomized selection of the remaining $n-k$ realizations so that these realizations are not completely ignored.

\section{Optimal Codeword Design under Randomized Selective Encoding}\label{sect:randomized}
The selective encoding scheme discussed so far is a deterministic scheme in which a fixed number of realizations are encoded into codewords and sent to the receiver node when realized. In this section, inspired by \cite{Mayekar18}, we consider a randomized selective encoding scheme. In this way, the most probable $k$ realizations are encoded with probability $1$. However, instead of discarding the remaining $n-k$ realizations, the source node encodes them with probability $\alpha$ and discards them with probability $1-\alpha$. In other words, in this model, less likely realizations that are not encoded under the highest $k$ selective encoding policy are sometimes transmitted to the receiver node. Thus, this randomized selective encoding policy strikes a balance between encoding every single realization and the highest $k$ selective encoding scheme discussed so far. In fact, under this operation, codewords for each one of the $n$ possible realizations need to be generated since every realization can be sent to the receiver node.

The transmitter node performs encoding using the following conditional probabilities
\begin{align}\label{cond_prob2}
   P_{X_\alpha}(x_i)= \begin{cases} 
      \frac{P_X(x_i)}{q_{k,\alpha}}, & i=1,2,\ldots, k \\
      \alpha \frac{P_X(x_i)}{q_{k,\alpha}}, & i=k+1,k+2,\ldots, n,   
   \end{cases}
\end{align}
where $q_{k,\alpha} \triangleq \sum_{\ell=1}^{k}P_X(x_\ell) +\alpha \sum_{\ell=k+1}^{n} P_X({x_\ell})$.

Next theorem determines the average age experienced by the receiver node under the randomized highest $k$ selective encoding scheme.

\begin{theorem}\label{thm_random}
    Under the randomized highest $k$ selective encoding scheme, the average age at the receiver node is given by
    \begin{align}
        \Delta_\alpha =  \frac{\mathbb{E}[L^2] + \frac{2}{q_{k,\alpha}\lambda} \mathbb{E}[L] + \frac{2}{(q_{k,\alpha}\lambda)^2}}{2\left(\mathbb{E}[L] +\frac{1}{q_{k,\alpha}\lambda}\right)}+\mathbb{E}[L], \label{avg_age5} 
    \end{align}
with $\mathbb{E}[L] = \sum_{i=1}^n P_{X_\alpha}(x_i) \ell(x_i)$, and  $\mathbb{E}[L^2] = \sum_{i=1}^n P_{X_\alpha}(x_i) \ell(x_i)^2$.
\end{theorem}{}
The proof of Theorem~\ref{thm_random} follows similarly to that of Theorem~\ref{thm1} by replacing $q_{k}$ with $q_{k,\alpha}$. 

Next, we formulate the optimization problem as,
\begin{align}
\label{problem3_mod}
\min_{\{ \ell(x_i), \alpha \}}  \quad &  \frac{\mathbb{E}[L^2]+2\bar{a}\mathbb{E}[L]+2\bar{a}^2}{2(\mathbb{E}[L]+\bar{a})} +\mathbb{E}[L] \nonumber \\
\mbox{s.t.} \quad & \sum_{i=1}^{n} 2^{-\ell(x_i)}\leq 1 \nonumber \\
\quad & \ell(x_i) \in \mathbb{R}^+, \quad i\in\{ 1,\ldots, n\},
\end{align}
where $\bar{a} = \frac{1}{\lambda q_{k,\alpha}}$. We first solve this problem for a fixed $\alpha$ in this section and determine the optimal $\alpha$ numerically for given arbitrary pmfs in Section~\ref{sect:num_res}. Following a similar solution technique as in Section~\ref{sect:opt_soln}, we find
\begin{align}\label{eqn:opt_lengths3}
\ell(x_i) =-\frac{\log \left( \frac{ P_{X_\alpha}(x_i)}{\beta (\log2)^2} W\left(\frac{\beta (\log2)^2}{P_{X_\alpha}(x_i)} 2^{\frac{-\theta +2\beta \log2+2\bar{a}}{3}} \right)\right)}{\log2},
\end{align} 
for $i\in \{1,2,\ldots, n\}$. To determine the age-optimal codeword lengths $\ell(x_i)$ for $i \in \{1, 2, \ldots, n\}$, we then employ the algorithm described in Section~\ref{sect:opt_soln}.

\section{Numerical Results} \label{sect:num_res}
In this section, we perform simulations to numerically characterize optimal $k$ values that minimize the average age for a given arbitrary pmf. For the first two simulations we use Zipf$(n,s)$ distribution with the following pmf for $n=100$, $s= 0.4$,
\begin{align}\label{zipf_pmf}
    P_X(x_i) = \frac{i^{-s}}{\sum_{j=1}^{n}j^{-s} }, \quad 1\leq i\leq N.
\end{align}

\begin{figure}[t]
	\centering  \includegraphics[width=0.98\columnwidth]{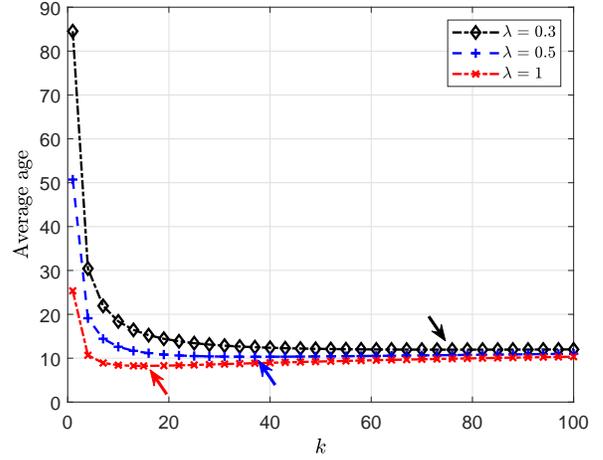}
	\caption{The average age values with the age-optimal codeword lengths for $\lambda \in \{0.3,0.5,1\}  $ for the pmf provided in (\ref{zipf_pmf}) with the parameters $n=100$, $s=0.4$. We vary $k$ from $1$ to $n$ and indicate $k$ that minimizes the average age for each $\lambda$ with an arrow.  }
	\label{sim1}
	\vspace{-4.5mm}
\end{figure}

In Fig.~\ref{sim1}, we show the effect of sending the most probable $k$ realizations when the update packets arrive at the transmitter node rather infrequently, i.e., the arrival rate is low. We consider the cases in which the arrival rate is equal to $\lambda = 0.3, 0.5,1$. For each arrival rate, we plot the average age as a function of $k=1,2,\ldots, n$. We see that increasing the arrival rate reduces the average age as expected. In this case, optimal $k$ is not equal to $1$ since the effective arrival rate is small. In other words, the transmitter node wants to encode more updates as opposed to idly waiting for the next arrival when the update arrivals are rather infrequent. Choosing $k$ close to $n$ is also not optimal as the service time for the status updates with low probabilities have longer service times which hurts the overall age performance. Indeed, in Fig.~\ref{sim1}, where update arrival rates are relatively small, it is optimal to choose $k=76$ for $\lambda = 0.3$, $k=37$ for $\lambda = 0.5$, and $k=15$ for $\lambda = 1$.        

\begin{figure}[t]
	\centering  \includegraphics[width=0.98\columnwidth]{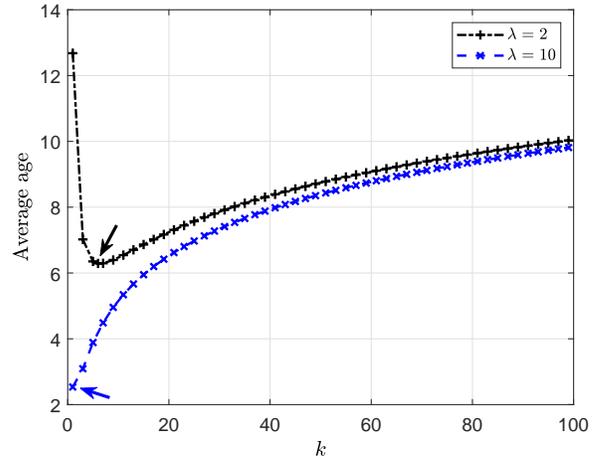}
	\caption{The average age values with the age-optimal codeword lengths for $\lambda \in \{2,10\}  $ for the pmf provided in (\ref{zipf_pmf}) with the parameters $n=100$, $s=0.4$. We vary $k$ from $1$ to $n$ and observe that choosing $k=1$ under the relatively high arrival rates ($\lambda =10$) minimizes the average age.}
	\label{sim2}
	\vspace{-5mm}
\end{figure}

In Fig.~\ref{sim2} we consider a similar setting as Fig.~\ref{sim1} but here update arrival rates are larger which means that updates arrive more frequently at the transmitter node. We observe that when $\lambda = 2$, the optimal $k$ is still not equal to $1$ (it is equal to 6 in Fig.~\ref{sim2}) as the updates are not frequent enough. However, once updates become more available to the transmitter node, i.e., the case with $\lambda=10$ in Fig.~\ref{sim2}, we observe that the transmitter node chooses to only encode the realization with the highest probability, i.e., $k=1$, and wait for the next update arrival instead of encoding more and more realizations which increases the overall codeword lengths thereby increasing the transmission times. We also observe that the average age decreases as the update arrival rate increases as in Fig.~\ref{sim1}.


\begin{figure}[t]
	\centering  \includegraphics[width=0.98\columnwidth]{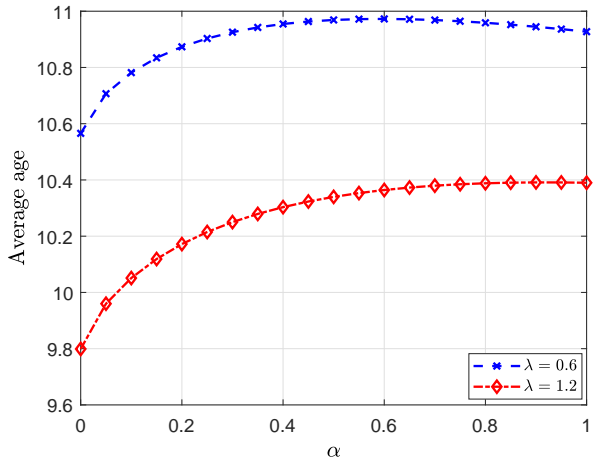}
	\caption{The average age values with the age-optimal codeword lengths for different $\alpha$ values with the pmf provided in (\ref{zipf_pmf}) with $n=100$, $s=0.2$ for $k=70$ when randomized highest $k$ selective encoding is implemented.}
	\label{sim5_randomized}
	\vspace{-0.5cm}
\end{figure}

We simulate the randomized highest $k$ selective encoding in Fig.~\ref{sim5_randomized} with Zipf distribution with parameters $n=100$, $s=0.2$. In Fig.~\ref{sim5_randomized}, we observe two different trends depending on the update arrival frequency at the source node, even though in either case, randomization results in a higher age at the receiver node than selective encoding, i.e., $\alpha=0$ case. When the arrival rate is high, $\lambda =1.2$ in Fig.~\ref{sim5_randomized}, we observe that age monotonically increases with $\alpha$ as randomization increases average codeword length, i.e., service times. Although increasing $\alpha$ results in a higher age at the receiver node, previously discarded $n-k$ realizations can be received under this randomized selective encoding policy. However, when the arrival rate is smaller, $\lambda =0.6$ in Fig.~\ref{sim5_randomized}, we observe that age initially increases with $\alpha$ and then starts to decrease because of the decreasing waiting times as opposed to increasing codeword lengths such that when $\alpha$ is larger than $0.3$, it is better to select $\alpha=1$, i.e., encoding every realization.

\section{Conclusions} \label{sect:conc}
We considered a status updating system in which an information source generates independent and identically distributed update packets based on an observed random variable $X$ which takes $n$ values based on a known pmf. The transmitter node implements a selective encoding policy to send the realizations to the receiver node such that only the most probable $k$ update realizations are encoded and the other realizations are discarded. We also considered a case in which the remaining previously discarded $n-k$ realizations are encoded into codewords randomly to further inform the receiver. We derived the average age for both types of operation and designed age-optimal codeword lengths. Through numerical results we showed that the proposed selective policy achieves a lower average age than encoding all the realizations and determined the age-optimal $k$ values for arbitrary pmfs. 

\bibliographystyle{unsrt}
\bibliography{IEEEabrv,lib_v5}

\begin{thebibliography}{10}

\bibitem{Kaul12a}
S.~K. Kaul, R.~D. Yates, and M.~Gruteser.
\newblock Real-time status: How often should one update?
\newblock In {\em IEEE Infocom}, March 2012.

\bibitem{Costa14}
M.~Costa, M.~Codrenau, and A.~Ephremides.
\newblock Age of information with packet management.
\newblock In {\em IEEE ISIT}, June 2014.

\bibitem{Bedewy16}
A.~M. Bedewy, Y.~Sun, and N.~B. Shroff.
\newblock Optimizing data freshness, throughput, and delay in multi-server
  information-update systems.
\newblock In {\em IEEE ISIT}, July 2016.

\bibitem{He16a}
Q.~He, D.~Yuan, and A.~Ephremides.
\newblock Optimizing freshness of information: On minimum age link scheduling
  in wireless systems.
\newblock In {\em IEEE WiOpt}, May 2016.

\bibitem{Sun17a}
Y.~Sun, E.~Uysal-Biyikoglu, R.~D. Yates, C.~E. Koksal, and N.~B. Shroff.
\newblock Update or wait: How to keep your data fresh.
\newblock {\em IEEE Transactions on Information Theory}, 63(11):7492--7508,
  November 2017.

\bibitem{Najm18b}
E.~Najm and E.~Telatar.
\newblock Status updates in a multi-stream {M/G/1/1} preemptive queue.
\newblock In {\em IEEE Infocom}, April 2018.

\bibitem{Najm17}
E.~Najm, R.~D. Yates, and E.~Soljanin.
\newblock Status updates through {M/G/1/1} queues with {HARQ}.
\newblock In {\em IEEE ISIT}, June 2017.

\bibitem{Soysal18}
A.~Soysal and S.~Ulukus.
\newblock Age of information in {G/G/1/1} systems.
\newblock In {\em Asilomar Conference}, November 2019.

\bibitem{Soysal19}
A.~Soysal and S.~Ulukus.
\newblock Age of information in {G/G/1/1} systems: Age expressions, bounds,
  special cases, and optimization.
\newblock May 2019.
\newblock Available on arXiv: 1905.13743.

\bibitem{Yates17b}
R.~D. Yates, P.~Ciblat, A.~Yener, and M.~Wigger.
\newblock Age-optimal constrained cache updating.
\newblock In {\em IEEE ISIT}, June 2017.

\bibitem{Hsu18b}
Y.~Hsu.
\newblock Age of information: Whittle index for scheduling stochastic arrivals.
\newblock In {\em IEEE ISIT}, June 2018.

\bibitem{Kadota18a}
I.~Kadota, A.~Sinha, E.~Uysal-Biyikoglu, R.~Singh, and E.~Modiano.
\newblock Scheduling policies for minimizing age of information in broadcast
  wireless networks.
\newblock {\em IEEE/ACM Transactions on Networking}, 26(6):2637--2650, December
  2018.

\bibitem{Gong19}
J.~Gong, Q.~Kuang, X.~Chen, and X.~Ma.
\newblock Reducing age-of-information for computation-intensive messages via
  packet replacement.
\newblock January 2019.
\newblock Available on arXiv: 1901.04654.

\bibitem{Buyukates19c}
B.~Buyukates and S.~Ulukus.
\newblock Timely distributed computation with stragglers.
\newblock October 2019.
\newblock Available on arXiv: 1910.03564.

\bibitem{Arafa19b}
A.~Arafa, K.~Banawan, K.~G. Seddik, and H.~V. Poor.
\newblock On timely channel coding with hybrid {ARQ}.
\newblock In {\em IEEE Globecom}, December 2019.

\bibitem{Sun17b}
Y.~Sun, Y.~Polyanskiy, and E.~Uysal-Biyikoglu.
\newblock Remote estimation of the {Wiener} process over a channel with random
  delay.
\newblock In {\em IEEE ISIT}, June 2017.

\bibitem{Sun18b}
Y.~Sun and B.~Cyr.
\newblock Information aging through queues: A mutual information perspective.
\newblock In {\em IEEE SPAWC}, June 2018.

\bibitem{Bastopcu19}
M.~Bastopcu and S.~Ulukus.
\newblock Age of information for updates with distortion.
\newblock In {\em IEEE ITW}, August 2019.

\bibitem{partial_updates}
D.~Ramirez, E.~Erkip, and H.~V. Poor.
\newblock Age of information with finite horizon and partial updates.
\newblock October 2019.
\newblock Available on arXiv:1910.00963.

\bibitem{Zou19b}
P.~Zou, O.~Ozel, and S.~Subramaniam.
\newblock Trading off computation with transmission in status update systems.
\newblock In {\em IEEE PIMRC}, September 2019.

\bibitem{Non_linear}
A.~Kosta, N.~Pappas, A.~Ephremides, and V.~Angelakis.
\newblock Age and value of information: Non-linear age case.
\newblock In {\em IEEE ISIT}, June 2017.

\bibitem{bastopcu_soft_updates_journal}
M.~Bastopcu and S.~Ulukus.
\newblock Minimizing age of information with soft updates.
\newblock {\em Journal of Communications and Networks}, 21(3):233--243, June
  2019.

\bibitem{Arafa17b}
A.~Arafa and S.~Ulukus.
\newblock Age minimization in energy harvesting communications:
  Energy-controlled delays.
\newblock In {\em Asilomar Conference}, October 2017.

\bibitem{Arafa17a}
A.~Arafa and S.~Ulukus.
\newblock Age-minimal transmission in energy harvesting two-hop networks.
\newblock In {\em IEEE Globecom}, December 2017.

\bibitem{Wu18}
X.~Wu, J.~Yang, and J.~Wu.
\newblock Optimal status update for age of information minimization with an
  energy harvesting source.
\newblock {\em IEEE Transactions on Green Communications and Networking},
  2(1):193--204, March 2018.

\bibitem{Arafa_Age_Online}
A.~Arafa, J.~Yang, and S.~Ulukus.
\newblock Age-minimal online policies for energy harvesting sensors with random
  battery recharges.
\newblock In {\em IEEE ICC}, May 2018.

\bibitem{Arafa18f}
A.~Arafa, J.~Yang, S.~Ulukus, and H.~V. Poor.
\newblock Online timely status updates with erasures for energy harvesting
  sensors.
\newblock In {\em Allerton Conference}, October 2018.

\bibitem{Arafa19e}
A.~Arafa, J.~Yang, S.~Ulukus, and H.~V. Poor.
\newblock Using erasure feedback for online timely updating with an energy
  harvesting sensor.
\newblock In {\em IEEE ISIT}, July 2019.

\bibitem{Farazi18}
S.~Farazi, A.~G. Klein, and D.~R. Brown~III.
\newblock Average age of information for status update systems with an energy
  harvesting server.
\newblock In {\em IEEE Infocom}, April 2018.

\bibitem{Yener_energy_19}
S.~Leng and A.~Yener.
\newblock Age of information minimization for an energy harvesting cognitive
  radio.
\newblock {\em IEEE Transactions on Cognitive Communications and Networking},
  5(2):427--439, May 2019.

\bibitem{Mayekar18}
P.~Mayekar, P.~Parag, and H.~Tyagi.
\newblock Optimal lossless source codes for timely updates.
\newblock In {\em IEEE ISIT}, June 2018.

\bibitem{Zhong16}
J.~Zhong and R.~D. Yates.
\newblock Timeliness in lossless block coding.
\newblock In {\em IEEE DCC}, March 2016.

\bibitem{Yates_Soljanin_source_coding}
J.~Zhong, R.~D. Yates, and E.~Soljanin.
\newblock Timely lossless source coding for randomly arriving symbols.
\newblock In {\em IEEE ITW}, November 2018.

\bibitem{Zhong17a}
J.~Zhong, E.~Soljanin, and R.~D. Yates.
\newblock Status updates through multicast networks.
\newblock In {\em Allerton Conference}, October 2017.

\bibitem{Zhong18b}
J.~Zhong, R.~D. Yates, and E.~Soljanin.
\newblock Multicast with prioritized delivery: How fresh is your data?
\newblock In {\em IEEE SPAWC}, June 2018.

\bibitem{Buyukates18}
B.~Buyukates, A.~Soysal, and S.~Ulukus.
\newblock Age of information in two-hop multicast networks.
\newblock In {\em Asilomar Conference}, October 2018.

\bibitem{Buyukates18b}
B.~Buyukates, A.~Soysal, and S.~Ulukus.
\newblock Age of information in multihop multicast networks.
\newblock {\em Journal of Communications and Networks}, 21(3):256--267, July
  2019.

\bibitem{Buyukates19}
B.~Buyukates, A.~Soysal, and S.~Ulukus.
\newblock Age of information in multicast networks with multiple update
  streams.
\newblock In {\em Asilomar Conference}, November 2019.

\bibitem{Cover}
T.~M. Cover and J.~A. Thomas.
\newblock {\em Elements of Information Theory}.
\newblock Wiley Press, 2012.

\bibitem{Yates14}
R.~D. Yates and D.~J. Goodman.
\newblock {\em Probability and Stochastic Processes}.
\newblock Wiley, 2014.

\bibitem{frac_programming}
W.~Dinkelbach.
\newblock On nonlinear fractional programming.
\newblock {\em Management Science}, 13(7):435--607, March 1967.

\bibitem{Boyd04}
S.~P. Boyd and L.~Vandenberghe.
\newblock {\em Convex Optimization}.
\newblock Cambridge University Press, 2004.

\bibitem{lambert}
R.~M. Corless, G.~H. Gonnet, D.~E.~G. Hare, D.~J. Jeffrey, and D.~E. Knuth.
\newblock On the {Lambert} {W} function.
\newblock {\em Advances in Computational Mathematics}, 5(1):329--359, December
  1996.

\end{thebibliography}
\end{document}